\documentclass[doublecol,linenumbers]{epl2} 
\usepackage{amssymb}
\usepackage{amsmath}
\usepackage{graphicx}
\usepackage{bm}
\usepackage{color}
\usepackage{subfigure}
\graphicspath{{./Figures/}}

\title{Distributed thermal tasks on many-body systems through a single quantum machine}
\shorttitle{Distributed thermal tasks on many-body systems through a single quantum machine}

\author{Bruno Leggio\inst{1} \and Pierre Doyeux\inst{1} \and Riccardo Messina\inst{1} \and Mauro Antezza\inst{1,2}}
\shortauthor{B. Leggio, P. Doyeux, R. Messina and M. Antezza}

\institute{
  \inst{1} Laboratoire Charles Coulomb (L2C), UMR 5221 CNRS-Universit\'{e} de Montpellier - F-34095 Montpellier, France\\
  \inst{2} Institut Universitaire de France, 1 rue Descartes, F-75231 Paris Cedex 05, France
}

\date{\today}

\abstract{We propose a configuration of a single three-level quantum emitter embedded in a non-equilibrium steady electromagnetic environment, able to stabilize and control the local temperatures of a target system it interacts with, consisting of a collection of coupled two-level systems. The temperatures are induced by dissipative processes only, without the need of further external couplings for each qubit. Moreover, by acting on a set of easily tunable geometric parameters, we demonstrate the possibility to manipulate and tune each qubit temperature independently over a remarkably broad range of values. These findings address one standard problem in quantum-scale thermodynamics, providing a way to induce a desired distribution of temperature among interacting qubits and to protect it from external noise sources.}

\begin{document}

\maketitle

\section{Introduction}
The study of thermodynamic phenomena at the quantum scale is nowadays becoming more and more challenging \cite{Horodecki2013,Scully2003,Mari2012,Skrzypczyk2014,Dorner2012,Leggio2013,Leggio2013b,Leggio2015b,Hovhannisyan2013,Eisert2015,Bellomo2012,Bellomo2012b}. This is certainly due to the rapid development of technologies capable of producing and addressing single quantum systems, which opened in these last years an unforeseeable range of possibilities to test and exploit quantum features at different scales \cite{Toyabe2010,Valenzuela2006,Todorov2010,Chen2012,Batalhao2014,Liu2014,Gessner2014}. One of the most investigated quantum thermodynamic systems is what goes under the name of absorption quantum thermal machine \cite{Linden2010,Levy2012,Leggio2015a,Venturelli2013,Correa2014}, under its many different forms. Studies of non-equilibrium or many-body quantum heat engines have been presented in \cite{Abah2014,Campisi2015}. Nevertheless, in the field of quantum absorption thermal machines a full account of the role of many-body interactions is still missing. Indeed, the theoretical description of their functioning has been mostly limited to models consisting of very few quantum systems, usually few two-level systems (2LSs or qubits) or three-level systems (3LSs), extracting heat from independent macroscopic reservoirs in order to perform thermodynamic tasks. Two regimes have received particular attention: the one under which a 3LS or few 2LSs can produce a steady heat flux between two macroscopic reservoirs \cite{Linden2010, Levy2012,Venturelli2013,Correa2014}, and the situation in which a quantum 3LS can tune the temperature of a single qubit \cite{Linden2010,Leggio2015a,Skrzypczyk2011,Brunner2014}, achieving either its refrigeration or its heating. Both these setups, although supplying important insight into quantum thermodynamics, cannot give a full account of the underlying atomic-scale thermodynamics, since the former does not allow any quantum correlations between the machine and its target, while the latter cannot account for collective phenomena in the target body. Thus the fundamental regime in which quantum effects are combined with many-body correlations in the underlying thermodynamic structure of machine-target interaction has not yet been explored.

Up to this level of description, the delivery of a thermodynamic task to a collection of $n_{\mathrm{q}}$ qubits would require the use of $n_{\mathrm{q}}$ machines. An even worse scenario is met when different local tasks must be delivered to a many-qubit system, the goal being to bring each qubit to a different temperature. In this case, in addition to $n_{\mathrm{q}}$ machines, also $n_{\mathrm{q}}$ sets of macroscopic reservoirs must be supplied, as their temperatures fix the steady task a machine can deliver.

A standard problem of this kind usually met in experiments is the necessity to keep interacting qubits at different temperatures in order for them to mediate and sustain steady energy flux from a colder to an hotter object, as theoretically proposed in \cite{Linden2010,Skrzypczyk2011,Venturelli2013}. However, practical realizations of this idea have been facing the nontrivial problem of keeping each qubit in thermal equilibrium with its corresponding object without allowing qubit-qubit thermalization \cite{Chen2012}. The need arises then for a simple and realistic scheme to induce a steady temperature distribution among many interacting 2LSs.

In this Letter we address this need, proving at the same time that $n_{\mathrm{q}}$ different temperatures on $n_{\mathrm{q}}$ qubits can be imposed and maintained through a \textit{single} 3LS system and a \textit{single} realistic out-of-thermal-equilibrium (OTE) electromagnetic environment.
As we will show, this \textit{multitasking quantum thermal machine} needs only two thermal baths and it can realize a broad variety of tasks. The temperatures induced by it on a set of qubits are easily tunable just by changing geometric parameters of the many-body target, even without modifying the temperature of the bath.

\section{Physical system}\label{physyst}
\begin{figure}[t!]
\begin{center}
\includegraphics[width=245pt]{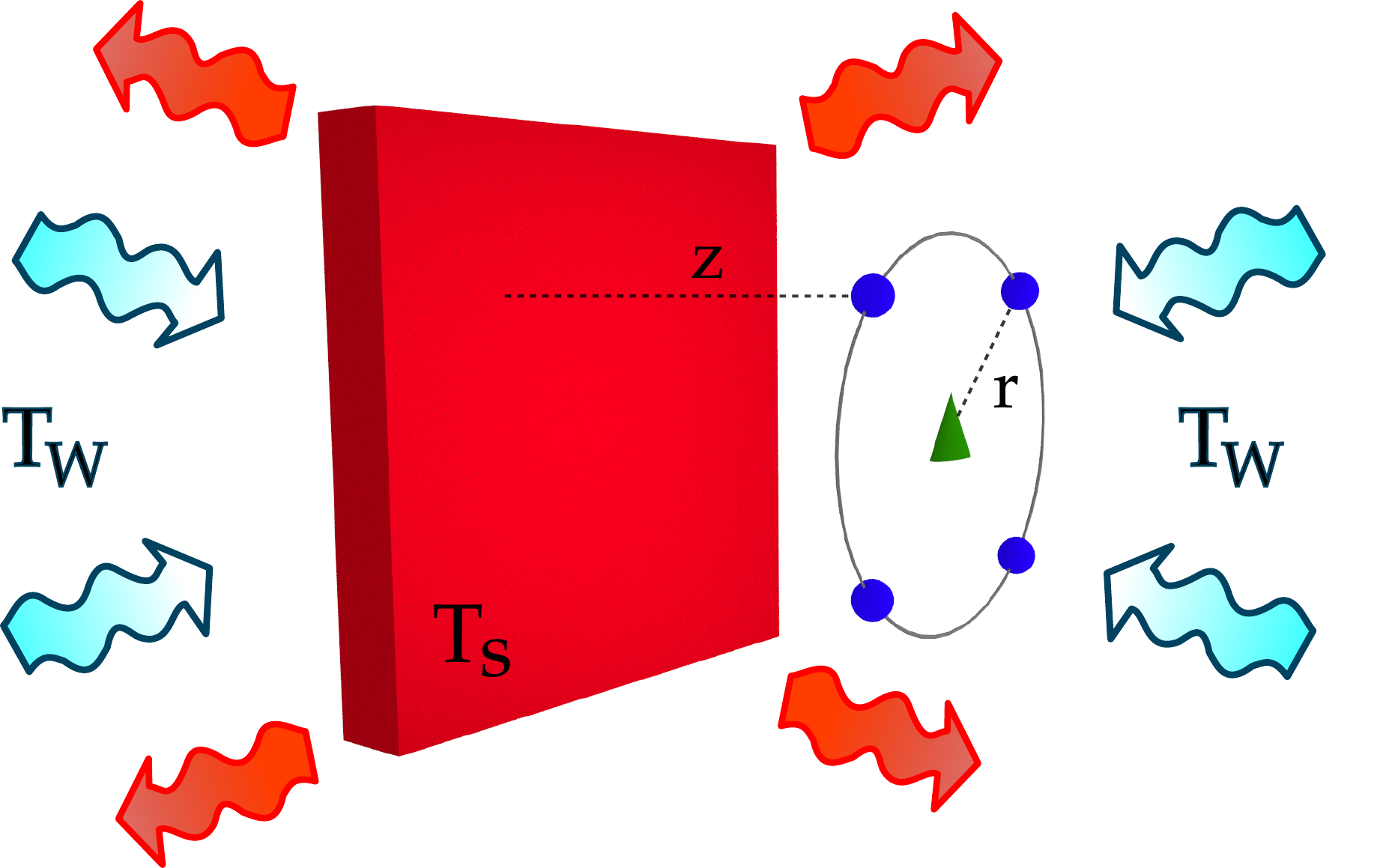}
\end{center}
\caption{The physical setup consists of an OTE electromagnetic field produced by a sapphire slab of thickness $\delta$, at temperature $T_S$, embedded in a thermal blackbody radiation at temperature $T_W\neq T_S$. Such a field plays the role of Markovian environment for a system of quantum emitters (atoms), all placed at the same distance $z$ from the slab surface: the qubit system is the target body, on which the three-level machine M (at distance $r$ from all the qubits) delivers thermodynamic tasks.}
\label{Figure1}
\end{figure}

Our multitasking machine is shown in Fig. \ref{Figure1}.
A sapphire slab of finite thickness $\delta=0.05\,\mu$m (whose optical data are taken from \cite{OpticalConstants}) is kept at a fixed temperature $T_S=900\,$K by means of a standard reservoir, and embedded in the blackbody radiation held at room temperature $T_W=300\,$K by the second thermal bath. The resulting electromagnetic field in the empty space thus neither is a blackbody radiation nor has a well defined temperature. Moreover, its spectral features are affected by the dielectric response function of the slab. At a distance $z$ from the slab a collection of quantum emitters (the target body TB and the 3LS machine M, hereby collectively referred to as atoms) is placed. In particular, the target body consists of a collection of $n_{\mathrm{q}}$ 2LSs, with all the qubits lying on a circle of radius $r$ and having the same transition frequency $\omega_a$. Finally, a 3LS is placed at the center of the circle. Its three transitions, all allowed, have different frequencies and one of the two low-frequency ones is in resonance with the qubits, while the high frequency transition is resonant with the first electronic resonance of sapphire at $\omega_S=0.81\times10^{14}\,$rad/s. We choose here $\omega_a=0.1\omega_S$. These parameters are known to produce the optimal working condition for the 3LS machine to act on a single qubit \cite{Leggio2015a}. The total Hamiltonian of the system is $H_{\mathrm{T}}=H_e+H_f+H_{ef}$, where $H_e$ is the free Hamiltonian of the quantum emitters, $H_f$ is the one of the electromagnetic field and $H_{ef}=-\sum_{n=1}^{n_{\mathrm{q}}}\mathbf{d}_n \cdot \mathbf{E}(\mathbf{R}_n)-\sum_{t=1}^3\mathbf{d}_{\mathrm{3LS}}^{(t)} \cdot \mathbf{E}(\mathbf{R}_{\mathrm{3LS}})$, describing the coupling between each emitter transition and the field $\mathbf{E}$ under the dipole approximation. Here $\mathbf{d}_n$, $n=1,\ldots,n_{\mathrm{q}}$ is the field-induced dipole of the $n$-th qubit at position $\mathbf{R}_n$, while $\mathbf{d}_{\mathrm{3LS}}^{(t)}$ is the field-induced dipole of the $t$-th transition of the 3LS at position $\mathbf{R}_{\mathrm{3LS}}$. All these dipoles are here assumed to have the same magnitude.

In the limit of weak emitters-field coupling \cite{BreuerBook}, the dynamics of the state $\rho$ of the sole atomic part of the system can be described by means of the Markovian master equation \cite{Bellomo2012,Bellomo2013,Bellomo2013b}
\begin{equation}\label{METQ}
\dot{\rho}=-\frac{i}{\hbar}\big[H_{\mathrm{at}},\rho\big]+D_{\mathrm{TB}}(\rho)+D_{\mathrm{M}}(\rho)+D_{\mathrm{nl}}(\rho),
\end{equation}
having introduced the effective atomic Hamiltonian $H_{\mathrm{at}}=H_e+H_{\mathrm{int}}$, consisting of the free part and of the dipole-dipole coupling between each pair of resonant atomic transitions, corresponding to exchange of virtual photons between two atoms.

For two resonant transitions $i$ and $j$, the coupling term has the form $h_{ij}=\Lambda_{ij}(\sigma^{\dag}_i\sigma_j+\sigma^{\dag}_j\sigma_i)$ and $H_{\mathrm{int}}=\sum_{i<j}h_{ij}$. Here $\sigma_i$ $(\sigma_i^{\dag})$ is the lowering (raising) operator of transition $i$ and $\Lambda_{ij}$ is a dipole-dipole coupling term whose expression in terms of the system parameters can be found in \cite{Bellomo2013}.

The dissipative part of the dynamics, accounted for by the term $D_{\mathrm{TB}}(\rho)+D_{\mathrm{M}}(\rho)+D_{\mathrm{nl}}(\rho)$, is described by superoperators of the general form
\begin{equation}
L(a,b)=a\rho b^{\dag}-\frac{1}{2}\big\{b^{\dag}a,\rho\big\}.
\end{equation}
The two dissipative terms $D_{\mathrm{TB}}(\rho)$ and $D_{\mathrm{M}}(\rho)$ characterize in particular local coupling of each transition in the target body and in the machine with the electromagnetic environment. They have the form of standard thermal-like dissipation, which for the $i$-th transition reads $D_i^{\mathrm{loc}}=\gamma_i^+D_i^{\mathrm{em}}(\rho)+\gamma_i^-D_i^{\mathrm{ab}}(\rho)$, where $D_i^{\mathrm{em}}(\rho)=L(\sigma_i,\sigma_i)$ and $D_i^{\mathrm{ab}}(\rho)=L(\sigma_i^{\dag},\sigma_i^{\dag})$ describe respectively emission and absorption of a photon through the transition $i$. Therefore $D_{\mathrm{TB}}(\rho)+D_{\mathrm{M}}(\rho)=\sum_iD_i^{\mathrm{loc}}$.

Finally, the dissipative term $D_{\mathrm{nl}}(\rho)$ couples resonant transitions, and is the real counterpart of the virtual process given by $H_{\mathrm{int}}$: it describes processes in which pairs of resonant transitions $i$ and $j$ collectively emit or absorb real photons into/from the field. For a fixed pair, this process is given by $D_{ij}^{\mathrm{nl}}=\gamma_{ij}^+D_{ij}^{\mathrm{em}}(\rho)+\gamma_{ij}^-D_{ij}^{\mathrm{ab}}(\rho)$, where $D_{ij}^{\mathrm{em}}(\rho)=L(\sigma_i,\sigma_j)$ and $D_{ij}^{\mathrm{ab}}(\rho)=L(\sigma_i^{\dag},\sigma_j^{\dag})$ have the same structure of $D_i^{\mathrm{em}}(\rho)$ and $D_i^{\mathrm{ab}}(\rho)$, but describe collective two-atom emission and absorption processes.

All these effects are a manifestation of emitters-field interaction, and their rates $\gamma_i^{\pm}, \gamma_{ij}^{\pm}$ and $\Lambda_{ij}$ are given in terms of the OTE field correlation functions, which at stationarity can be exactly calculated \cite{Messina2011,Messina2011b,Messina2014}. They depend on the transition frequency, the interatomic separations, the atoms-slab distance, the dipole orientations and the slab dielectric properties, geometry and temperature. The detailed expressions for all these parameters are derived and discussed in \cite{Bellomo2013}.

Two transitions at different frequencies, all the other parameters being the same, have thus different photon emission/absorption rates. In other words, they feel their respective local environments at different temperatures, the associated Boltzmann factor at the transition frequency being $\gamma_i^-/\gamma_i^+$. As a consequence, the 3LS perceives three different field temperatures on its three transitions, such that its populations do not follow a simple thermal distribution \cite{Bellomo2012,Leggio2015a}. Referring to the three levels of M as $|0\rangle, |1\rangle$ and $|2\rangle$, if the transition $|0\rangle \leftrightarrow|1\rangle$ is in contact with a colder effective environment than the transition $|0\rangle \leftrightarrow|2\rangle$, the dynamics thus induced tends to increase the ratio $p_2/p_0$ and to reduce $p_1/p_0$. As such, this process has the net effect of increasing the ratio $p_2/p_1$ which is equivalent to increasing the temperature $\theta_\text{M}$ of the transition $|1\rangle \leftrightarrow|2\rangle$ (defined through its Boltzmann factor $p_2/p_1=\exp{\big[-\hbar \omega_{12}/(k_B \theta_\text{M})\big]}$).

The OTE field is thus an optimal tool to manipulate the temperatures of atomic transitions in a 3LS. If, in turn, the 3LS is coupled to a qubit through the resonant dipole-dipole interaction $\Lambda$ on its transition $|1\rangle \leftrightarrow|2\rangle$, such a 2LS will thermalize to a temperature $\theta_{\mathrm{2LS}}$ intermediate between $\theta_\text{M}$ and the one of its effective environment (hereby referred to as $T_{\mathrm{env}}$) \cite{Leggio2015a}. Since $T_{\mathrm{env}}$ is the natural temperature at which the qubit would thermalize in the absence of M, any difference between $\theta_{\mathrm{2LS}}$ and $T_{\mathrm{env}}$ signals the achievement of a thermal task by the 3LS on the 2LS. In particular, when $\Lambda\gg \gamma_{\mathrm{2LS}}^+$ (which is the case for a broad range of machine-qubit distances \cite{Bellomo2013}), the qubit temperature $\theta_{\mathrm{2LS}}$ will be steadily driven very close to the one of M. Through this mechanism, it has been shown in \cite{Leggio2015a} that a qubit can be refrigerated, heated up and even brought to population inversion by the same field-machine configuration, just by changing the atoms-slab distance $z$.

The purpose of this Letter is to exploit this mechanism to induce a steady distribution of temperatures $\theta_n$ on $n_{\mathrm{q}}$ qubits, with in general $\theta_n\neq \theta_m$ for $n\neq m$ in $1,\ldots,n_{\mathrm{q}}$.

\section{Machine-induced thermodynamics}\label{thermodyn}
\begin{figure}[t!]
\begin{center}
\includegraphics[width=150pt]{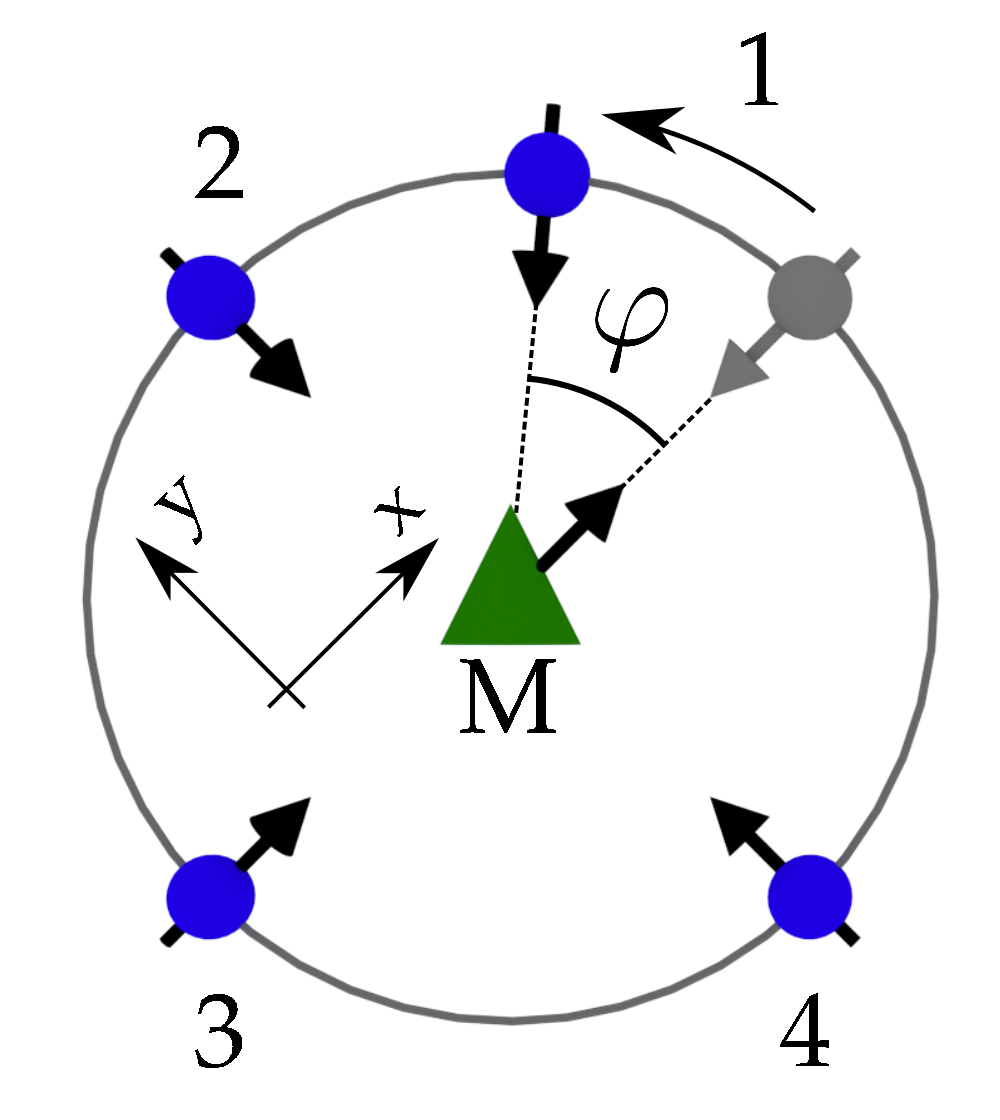}
\end{center}
\caption{Model system. Four 2LSs are placed along a circle around the 3LS M, with their dipoles pointing always toward M. The grey configuration for qubit 1 corresponds to the regular disposition of qubits, and it defines the value $\varphi=0$. Such a configuration also defines the reference frame of the system, with the $x$ and $y$ axes oriented as in the figure.}
\label{Figure2}
\end{figure}

In tackling this question, we analyze a model system where the target body is composed of $n_{\mathrm{q}}=4$ qubits. Initially each 2LS is placed on one vertex of a square inscribed in the circle on which the qubits lie, which defines the $xy$ plane as shown in Fig. \ref{Figure2}. Their dipoles, all of the same magnitude, point toward the center of the circle where the machine M is placed. The dipole of M points toward one 2LS, labeled as qubit 1. This corresponds to the configuration shown in Fig. \ref{Figure2} with $\varphi=0$. All the results shown here are obtained with the open-source package QuTiP \cite{qutip}.

\begin{figure}[t!]
\begin{center}
\includegraphics[width=245pt]{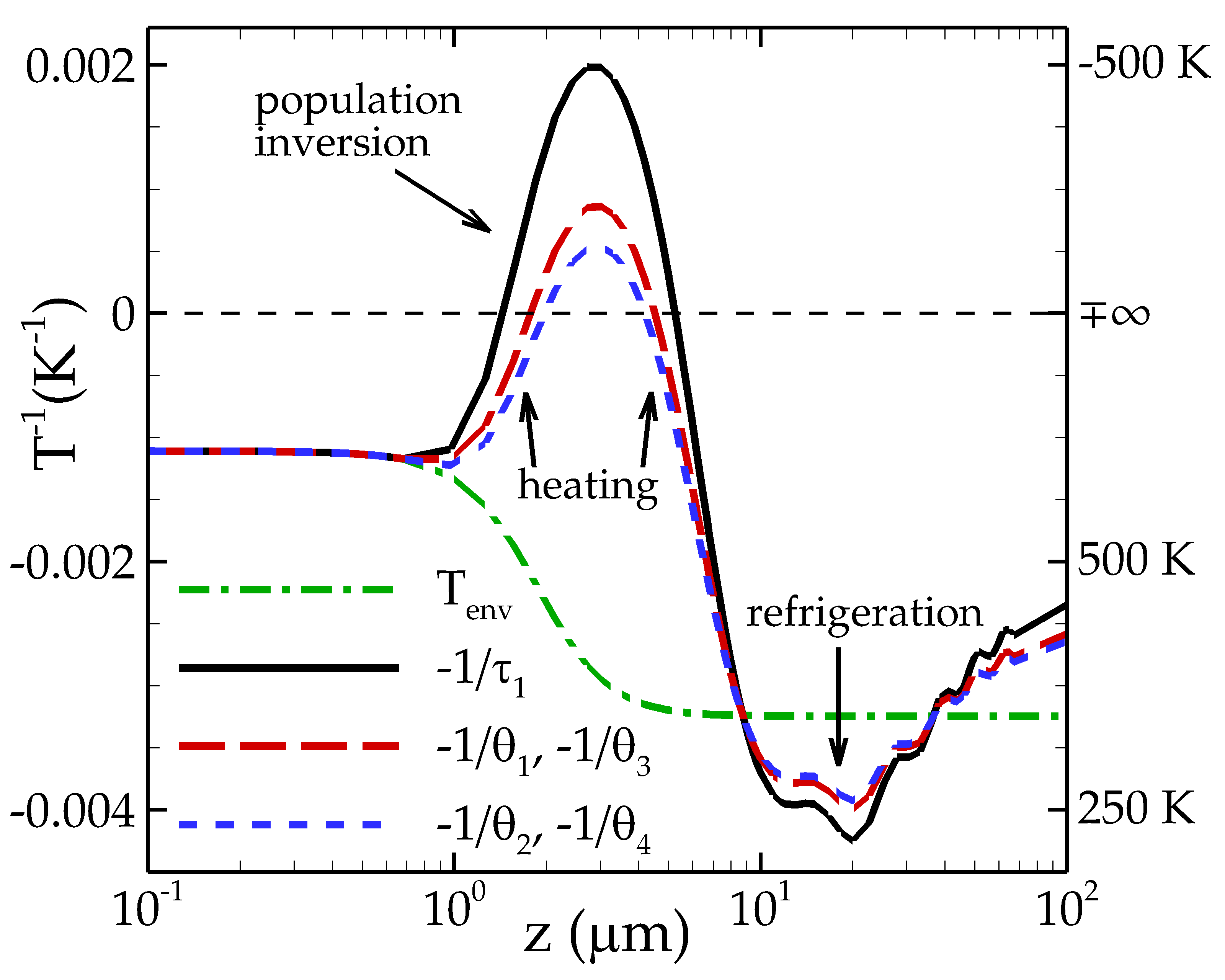}
\end{center}
\caption{Steady qubits temperatures, represented as $-1/\theta_k$ for $k=1,3$ (long-dashed red line), for $k=2,4$ (short-dashed blue line), $-1/\tau_1$ for the 1-qubit case (solid black line), compared with the qubits environmental temperature $T_{\mathrm{env}}$ (dot-dashed green line), versus the atoms-slab distance $z$. The qubits are in a regular disposition, being placed on the vertices of a square, with their dipoles pointing toward the circumcenter of the square. In both the 1-qubit and the 4-qubit cases, the 2LS-machine distance is $r=75\,\mu$m.}
\label{Figure3}
\end{figure}

The pair $\{\text{M}, 1\}$ is thus in a configuration analogous to the one studied in \cite{Leggio2015a}. Fig. \ref{Figure3} gives the steady temperatures $\theta_k$, $k=1,\ldots,4$ of each qubits versus $z$ for $r=75\,\mu$m, compared with the steady temperature $\tau_1$ reached by the 2LS in the 1-qubit case with the same qubit-machine distance. For graphical purposes, Fig. \ref{Figure3} shows the behaviour of $-1/\theta_k$, $-1/\tau_1$ and $-1/T_{\mathrm{env}}$, with the correspondent values of temperature given on the right vertical scale. The temperature of each qubit induced by their interaction with M follows the same qualitative $z$-behaviour and the same tasks as the one achieved for the single qubit case. Thus, despite a quantitative change between the 1- and the 4-qubit configurations ($\sim 70\%$ for the peak in population inversion, $\sim 0.6\%$ for the peak in refrigeration), this thermodynamic setup can deliver tasks also on many-qubit systems. However, as one sees from Fig. \ref{Figure3}, all the qubits temperatures are almost the same, thus no significative steady distribution can be achieved in this configuration. For instance, note that in this regular disposition of qubits, $\theta_1=\theta_3$ and $\theta_2=\theta_4$, all of them being anyway much closer to each other than to $T_{\mathrm{env}}$.

Before proceeding further, note that the regular 2LSs disposition just considered is quite a particular one, being characterized by the fact that the dipoles of qubits 2 and 4 are orthogonal to the one of M, all of them being parallel to the slab surface. In this condition, neither the dipole-dipole coupling $\Lambda$ nor the collective dissipation $\gamma_{ij}$ couples M to 2 and 4 \cite{Bellomo2013}. This means that qubits 2 and 4 do not perceive at all the presence of M. Their temperature change with respect to $T_{\mathrm{env}}$ is thus due to their interaction with the rest of the qubits system, which on the one hand undergoes the effect of M and, on the other, relays this effect on the pair $\{2,4\}$. This is an example of very effective interplay among dissipative effects, atom-atom pairing (the interaction between M and each atom of the pair $\{1,3\}$) and many-body interactions within the target body. As a consequence, we argue that by tuning one of these two features one can exert a control on the way the machine task is distributed among the 2LSs. In particular, we will consider two control parameters in what follows: the geometrical position of atom 1 along the circle circumscribing the square and the direction of the dipole of atom 4. Both of these parameters will affect one or both of the two fundamental interactions on which the task distribution is based. In particular, moving one qubit along the circle will change its coupling strength with both the other 2LSs and M, introducing a bias in the many-body interactions within the target body and in the machine-body coupling. The orientation of the dipole of an appropriately chosen 2LS, on the other hand, will only affect its interaction with the rest of the qubits.

\section{External control of the steady temperature distribution}
Consider thus the atomic system to be placed at a distance $z\simeq 2.7\,\mu$m from the slab, where the peak of population inversion is achieved for a single qubit. We now begin to change the position of qubit 1, by rotating it of an angle $\varphi$ around the point where the machine is placed, as shown in Fig. \ref{Figure2}. Whereas its dipole always points toward M, its interaction with the rest of the qubits will significantly change, growing much stronger as qubit 1 approaches one of the other 2LSs. For each angle $\varphi$, we imagine to wait long enough for the system to reach a steady state, referred to as $\rho(\varphi)$ and depending on the angular position $\varphi\in [0,2\pi]$ of qubit 1. All the thermodynamic quantities evaluated on this state will then be functions of $\varphi$. $\varphi$ is defined such that $\varphi=(k-1)\pi/2$ corresponds to the regular position of qubit $k=1,2,3,4$ on the square. Of course, the three positions $\varphi=\pi/2,\pi,3\pi/2$, where qubit 1 would sit on top of another 2LS, are idealized situations, and correspond to infinite interaction strength $\Lambda$. These values of $\varphi$ must then be read as limiting cases for two qubits being much closer to each other than to any other constituent of the atomic system.

\begin{figure}[t!]
\begin{center}
\includegraphics[width=245pt]{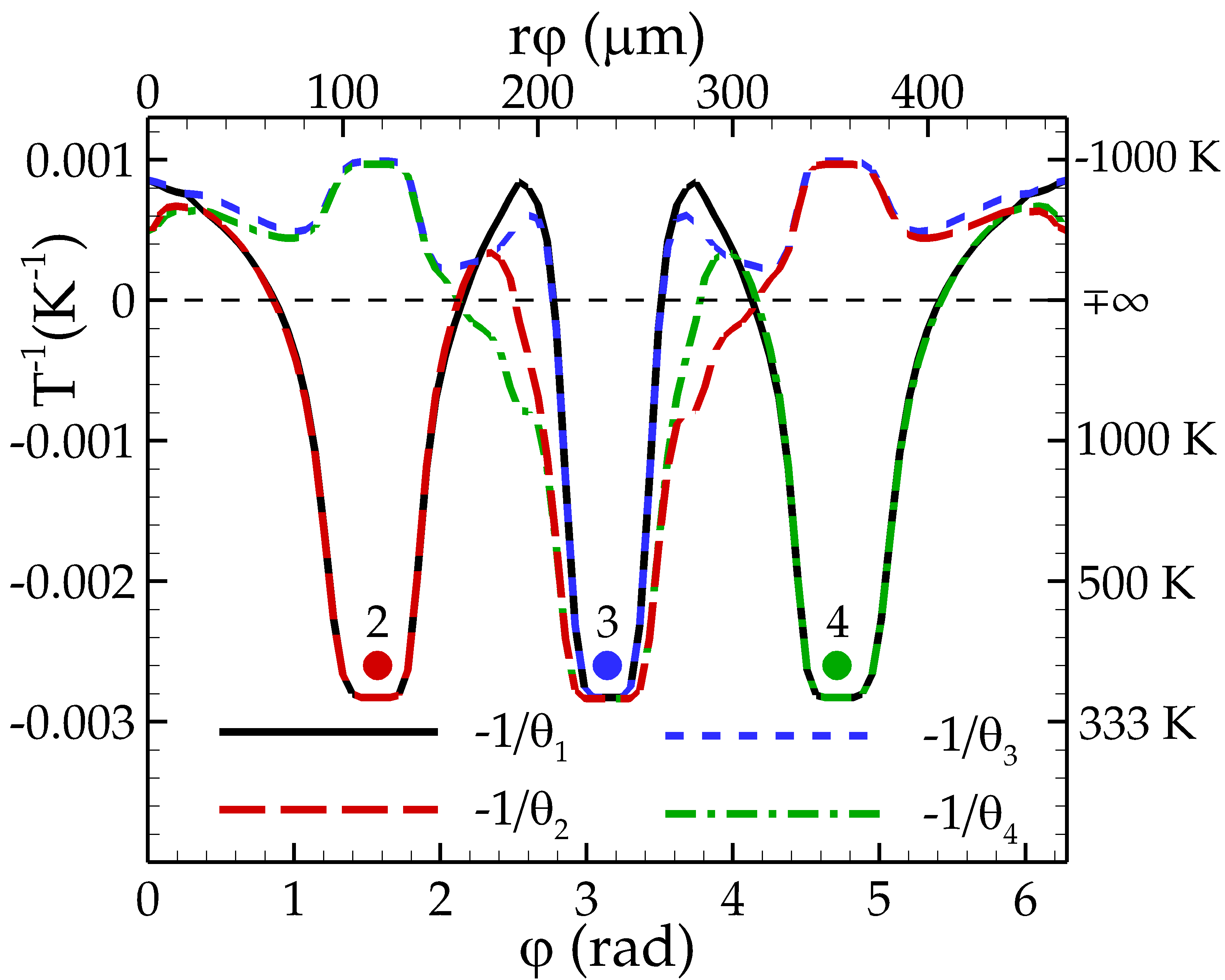}
\end{center}
\caption{Steady qubits temperatures represented as $-1/\theta_k$ for $k=1$ (solid blue line), 2 (long-dashed red line), 3 (short-dashed blue line) and 4 (dot-dashed green line), versus the angular displacement $\varphi$ of qubit 1. The other geometrical parameters are $z\simeq 2.7\,\mu$m and $r=75\,\mu$m. The plateau regions correspond to the approach of one of the values $\varphi=(k-1)\pi/2$ for $k=2,3,4$ (the atomic positions are represented by colored circles) and describe the situation of very strong coupling between two qubits. The upper horizontal scale reports the values of the linear distance $r \varphi$ travelled by qubit 1 along the circle.}
\label{Figure4}
\end{figure}

Fig. \ref{Figure4} reports the steady temperatures of each 2LS as a function of $\varphi$. As one can see, the temperature of qubit 1 gets strongly modulated by $\varphi$, repeatedly crossing the population inversion line. Moreover, depending on its angular displacement, qubit 1 interacts very strongly with one of the other 2LSs, such that the pair is effectively decoupled from the rest of the atomic system. In these cases, the temperatures of qubit 1 and of its paired 2LS fall down to the value of $T_{\mathrm{env}}$ (the lower plateau around 350 K). This is due to the fact that very strong correlations between qubit 1 and its paired 2LS are built due to the vicinity of the 2 atoms, a situation in which monogamy forbids the pair to create any kind of correlations with anything external. As such, the paired atoms are completely decoupled from the machine and, as a consequence, equilibrate with their local environment. The width of these plateaus corresponds to the regime in which the dipole-dipole interaction amplitude $\Lambda_{1j}$ between the paired atoms is much larger than any other coupling strength of the system. Note indeed that, in a broad range of $\varphi$ before and after such plateaus, the temperatures of the two paired qubits are almost overimposed. The central plateau at $\varphi=\pi$ is special, as it corresponds to a complete decoupling of the target body from the machine. This is due to the dipole orientation chosen here: 1 and 3 are strongly paired and as such decoupled from M, while the two 2LSs left do not directly interact with M whatsoever due to their dipoles being orthogonal to $\mathbf{d}_{\mathrm{M}}$. As such, the machine is unable to deliver any task on TB, which thermalizes as a whole with the environment at $T_{\mathrm{env}}$.

The regions around plateaus provide thus a mechanism to significantly divide the 2LSs temperatures into two groups, one at positive and one at negative values in the example analyzed here (and, more in general, one close to $\theta_\text{M}$ and one close to $T_{\mathrm{env}}$). But probably the most interesting region for the purpose of creating a steady temperature distribution in TB is the one in between two consecutive plateaus. For $\pi/2<\varphi<\pi$, for instance, the four qubits temperatures are all different, the same happening for $\pi<\varphi<3\pi/2$. Given the radius of the circle on which qubit 1 moves, the linear difference between the two positions at $\varphi=\pi/2$ and $\varphi=\pi$ is approximately of $106\,\mu$m (or $117\,\mu$m along the circle), which is a realistic and achievable scale of distances between two quantum emitters in real experiments \cite{Whitlock2009}. Within this scale, a broad distribution of steady qubit temperatures can be realized: consider the exemplary case of $\varphi=2.5\,$rad: the four qubits temperatures are $\theta_1=-1190\,$K, $\theta_2=6700\,$K, $\theta_3=-1820\,$K and $\theta_4=1280\,$K.

We stress here that any dissipative dynamics inducing different states on single constituents of a many-qubit system creates a distribution of local temperatures, since a 2LS is always in a Gibbs state. The novelty here is the achievement of a detailed control over such a distribution only by means of geometrical parameters. In particular, by appropriately modulating $z$, $\varphi$ and possibly $r$ one can exert such control over local thermal state of each constituent of the multipartite target body. Such an effect is also very robust against interaction with thermal reservoirs in the weak-coupling limit, as already signaled by the difference between $\theta_k$ of each qubits and $T_{\mathrm{env}}$ of their electromagnetic bath. One can thus engineer proper values of local qubit temperatures in order to assure the proper functioning of them as, for instance, thermal machine for macroscopic bodies \cite{Linden2010}. Note that the physics just described remains the same when analyzed at different values of $z$, the only difference being in the range of values for the various $\theta$. Such a range is indeed fixed to be a subinterval of $[T_{\mathrm{env}},\theta_\text{M}]$ (or of $[\theta_\text{M},T_{\mathrm{env}}]$ in the case of refrigeration).

\begin{figure}[t!]
\begin{center}
\includegraphics[width=245pt]{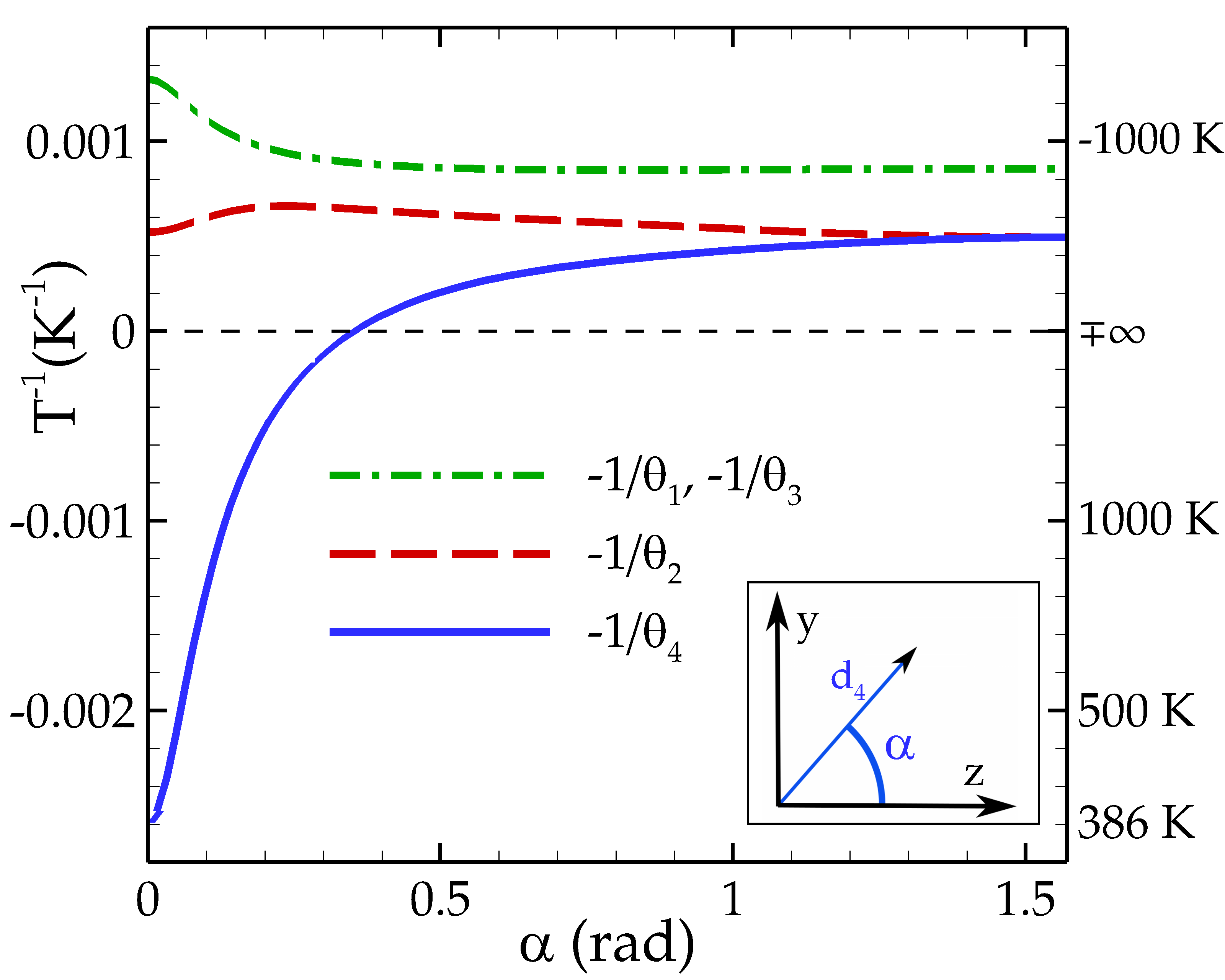}
\end{center}
\caption{Steady qubits temperatures represented as $-1/\theta_k$ for $k=1,3$ (dot-dashed green line), 2 (long-dashed red line) and 4 (solid black line), versus the orientation $\alpha$ of the dipole of qubit 4. The other geometrical parameters are $z\simeq 2.7\,\mu$m and $r=75\,\mu$m.}
\label{Figure5}
\end{figure}

The second parameter whose tuning we want to explore is the direction of one of the qubits dipoles, as shown in the inset of Fig. \ref{Figure5}. We choose here in particular the 2LS we refer to as qubit 4 in Fig. \ref{Figure2}. Its feature is to have its dipole orthogonal to the one of M, and as such not to interact directly with it. As a consequence, a rotation of its dipole in the plane $yz$, as represented in the inset of Fig. \ref{Figure5}, keeps the 2LS always decoupled from the machine and only affects its interaction with the rest of the qubits. Fig. \ref{Figure5} shows the dependence of the qubits temperatures on the angle $\alpha$ between the direction of the dipole of qubit 4 and the $z$ axis. As such, $\alpha=0$ means a dipole orthogonal to the atomic plane and thus to the slab surface, while $\alpha=\pi/2$ corresponds to the dipole pointing toward the machine.

The possibility of controlling the value of $\theta_4$ is evident here, the temperature ranging continuously from $386\,$K for $\alpha=0$ to $-2022\,$K for $\alpha=\pi/2$. The values of the other three qubits temperatures are only slightly affected by $\alpha$ and are relatively close to each other. As such, the rotation of one dipole is an efficient mechanism to decouple the temperature of one qubit from the rest of the system, and to assign it a steady value at will. The underlying mechanism must of course be looked for in the structure of dipole-dipole coupling: for $\alpha=0$, qubit 4 is decoupled from the rest of the atomic system \footnote{As a matter of fact, having fixed a particular pair of atoms and calling $\hat{w}$ the direction joining two atoms and $\hat{h}$ the one perpendicular to the atomic plane, a $wh$ interaction between dipoles exists, due to the electromagnetic contribution of the slab \cite{Bellomo2013}. In practice, however, such an interaction is negligibly small with respect to co-planar dipoles couplings and will thus have no detectable consequence on the qubits thermodynamics.} and as such thermalizes with its local environment at the temperature $T_{\mathrm{env}}=386\,$K. By increasing $\alpha$ one triggers and strengthens the coupling of the 2LS labeled as 4 with the rest of the target body, and starts receiving then, indirectly, the task of the machine. As long as the qubit-qubit coupling is comparable with the coupling of qubit 4 with its local environment, however, the steady value of $\theta_4$ will be something in between $T_{\mathrm{env}}$ and $\theta_\text{M}$. This competition between the driving of the local field and the task of the machine varies continuously and smoothly with $\alpha$, allowing one to adjust it at will. Finally, for $\alpha$ big enough, the machine effect becomes dominant and $\theta_4$ becomes almost independent on the dipole orientation of qubit 4.

\section{Conclusions}
In this Letter, we have demonstrated the possibility of externally control and distribute the temperature of each constituent of a many-qubit quantum system. The mechanism we propose is simple and yet powerful and stable: by means of a single three level system and an out-of-thermal-equilibrium steady electromagnetic field one can tune and stabilize the local qubit temperatures to values at will in a remarkably broad range, only depending on the details of the three level system and of the realistic parameters characterizing the electromagnetic field. The local temperatures can be easily tuned, by means of easily handleable parameters such as qubits position or dipole orientations. The OTE field naturally mediates for the interaction between qubits and between them and the 3LS. Moreover, the latter is usually much stronger than the one between each qubit and a possible thermal reservoir connected to it, such that our paradigm overcomes the usual problem of obtaining a configuration of interacting qubits having however strongly different temperatures. This need arises when practically realizing microscopic machines to produce steady heat fluxes between meso- or macroscopic reservoirs \cite{Linden2010}, for which each of the interacting qubits must also be in local thermal equilibrium with one of these reservoirs.

Our results thus pave the way for the practical exploitation of the new generation of microscopic and quantum thermal machines recently thoroughly studied.

\begin{acknowledgements}
We acknowledge financial support from the Julian Schwinger Foundation.
\end{acknowledgements}

\end{document}